\documentclass[%
reprint,
 amsmath,amssymb,
 aps,
]{revtex4-1}
\usepackage{epsfig,epsf,rotating}
\usepackage{dcolumn}
\usepackage{bm}
\usepackage{graphicx}

\begin{document}

\preprint{APS/123-QED}

\title{Anomalous behavior of pion production in high energy particle collisions}

\author{\firstname{A.~A.}~\surname{Bylinkin}}
 \email{alexander.bylinkin@desy.de}
\affiliation{%
 Institute for Theoretical and Experimental
Physics, ITEP, \\
Moscow, Russia
}%
\author{\firstname{A.~A.}~\surname{Rostovtsev}}
 \email{rostov@itep.ru}
\affiliation{%
 Institute for Theoretical and Experimental
Physics, ITEP, \\
Moscow, Russia
}%

\date{\today}

\begin{abstract}
A shape of invariant differential cross section for charged hadron
production as function of transverse momentum measured in various collider experiments  is analyzed. Contrary to the behavior of produced charged kaons, protons and antiprotons the pion spectra require an anomalously high contribution of an exponential term to describe the shape. 
\end{abstract}

\pacs{Valid PACS appear here}
\maketitle


    There exists a large body of experimental data on hadron production in high energy particle collisions. A systematic comparative analysis of the vast volume of available experimental data allows to gain an insight into the hadron production mechanism. In the present paper the production of charged kaons, protons and antiprotons is analyzed within a framework of the approach developed in our previous publication~\cite{OUR1}.

In~\cite{OUR1} a new parameterization of the spectrum shape of charged pions inclusively produced in the central rapidity region in high energy particle collisions was proposed. 
According to~\cite{OUR1} the pion spectra are approximated using a modified Tsallis-type function~\cite{Tsallis}:
\begin{equation}
\label{eq:exppl}
\frac{d\sigma}{P_T d P_T} = A_e\exp {(-E_{Tkin}/T_e)} +
\frac{A}{(1+\frac{P_T^2}{T^{2}\cdot n})^n},
\end{equation}
where  $P_T$ is transverse momentum of the produced particle, $E_{Tkin} =
\sqrt{P_T^2 + M^2} - M$
with M equal to the produced hadron mass. $A_e, A, T_e, T$ and "n" are the free parameters to be determined by a fit to the data.
For the reasons for choice of this particular parameterization~(\ref{eq:exppl}) see detailed discussion in~\cite{OUR1}.

\begin{figure}[h]
\includegraphics[width =8cm]{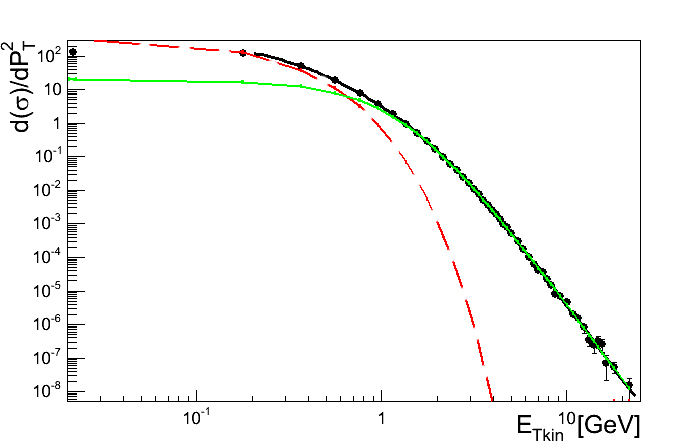}
\caption{\label{fig:01} Pion spectrum~\cite{UA1} fitted with a modified Tsallis function~(\ref{eq:exppl}): the red (dashed) line shows the exponential term and the green (solid) one - the power law.}
\end{figure}

The proposed new parameterization is represented by a sum of an exponential and a power law functional terms. The variations of the parameters of this approximation were studied as function of energy and type of colliding particles, as well as of other experimental conditions.
A typical charged particle spectrum as function of transverse energy, fitted with this function~(\ref{eq:exppl}) is shown in Fig~\ref{fig:01}. 

The contributions of the exponential and power law
terms
of the parameterization~(\ref{eq:exppl}) to
the typical spectrum of charged particles produced in $pp-$collisions are also shown separately in Fig~\ref{fig:01}.
The relative contribution of these terms is characterized by ratio $R$ of the power law term alone to the parameterization function integrated over $P_T^{2}$:
\begin{equation}
R = \frac{AnT}{AnT + A_e(2MT_e + 2T_e^2)(n-1)}
\end{equation}
One can notice that for charged particle (mainly charged pions) production the exponential term dominates. Moreover, it was shown~\cite{OUR1} that this ratio $R$ has only weak dependence (if any) on the collision energy in $pp-$interactions.

The present analysis of produced charged kaon, proton and antiproton spectra  is based
on the published hadron production data from
$pp-$collisions
at RHIC~\cite{Adare:2011vy} and LHC~\cite{Aamodt:2011zj} and
$AuAu-$collisions with
different centralities at RHIC~\cite{RHIC,Adcox:2001mf,STAR}.
The data for
all these inclusive differential cross section measurements have been
taken with a minimum
bias trigger conditions and at
center of mass energy $(\sqrt{s})$ ranging from 63 to 900 GeV.
The results of the analysis are discussed along with those for the pion spectra measured in different experiments~\cite{ISR,UA1,CDF,H1gp,OPAL,RHIC,CMS} and reported in~\cite{OUR1}.

\begin{figure*}[!]
\includegraphics[width = 18cm]{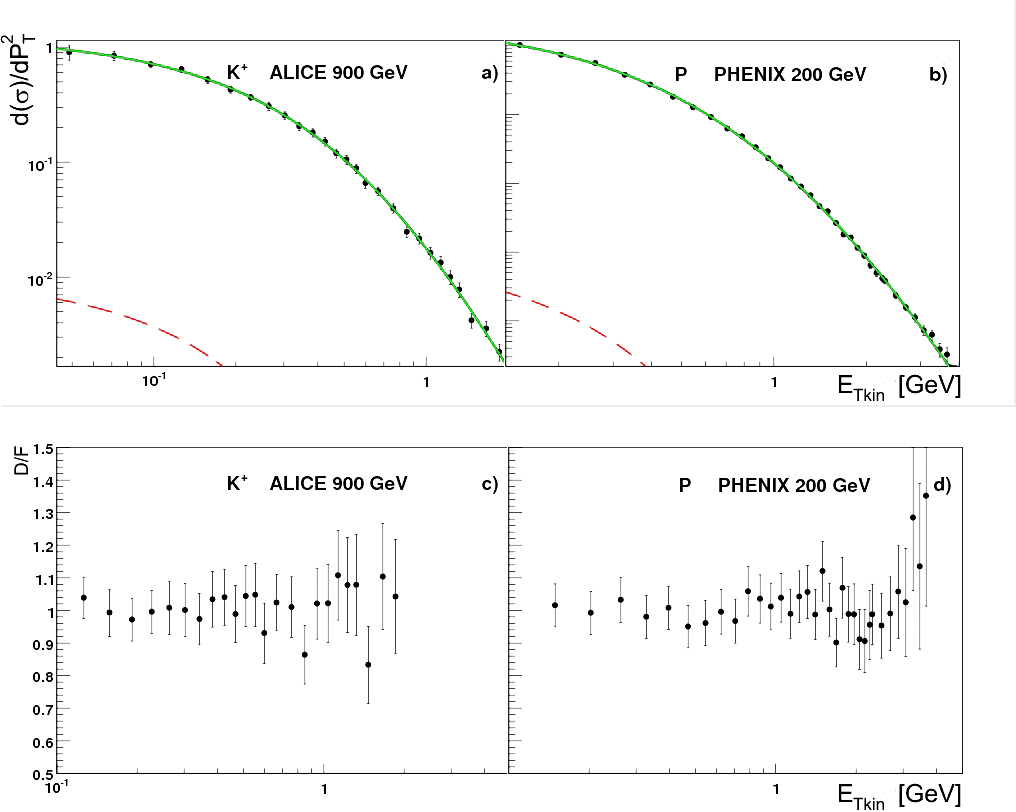}
\caption{\label{fig:02} Kaon and Proton spectra fitted with function~(\ref{eq:exppl}): the red (dashed) line shows the exponential term and the green (solid) one - the power law, and corresponding data-to-fit ratios.}
\end{figure*}

\begin{figure}[h]
\includegraphics[width = 8cm]{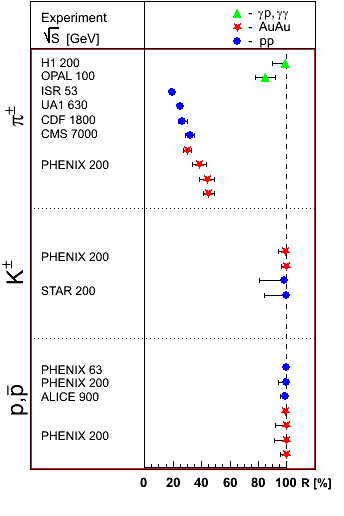}
\caption{\label{fig:03} Power law term contribution $R$ for pions, kaons and protons produced in various collider experiments.}
\end{figure}

Contrary to the pions, the spectra of charged kaons and protons produced in
high energy $pp$ and $AuAu$ collisions demonstrate a quite different behavior. Fitting these spectra to the parameterization function~(\ref{eq:exppl}) (Fig~\ref{fig:02}a,b) shows no room for the contribution of the exponential term. The data-to-fit ratios ($D/F$) on Fig~\ref{fig:02}c,d prove that the power law term alone provides an excellent description of the kaon and proton spectra. The values of the ratio $R$ obtained for different hadrons produced in various types of high energy interactions are plotted together in Fig~\ref{fig:03}. This figure clearly shows that the sizable exponential term exists only in the pion spectra produced in $pp$ and heavy ion collisions. 
\begin{figure}[h]
\includegraphics[width = 8cm]{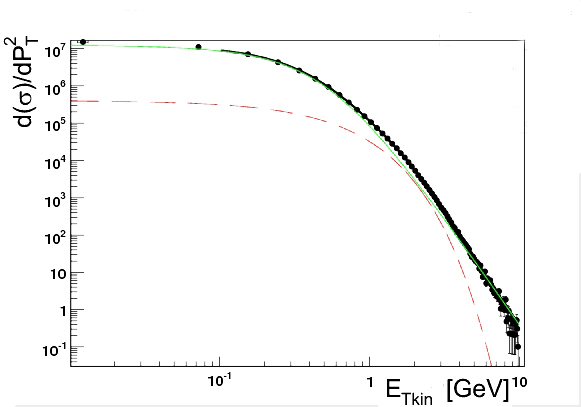}
\caption{\label{fig:04} MC generated spectrum fitted with function ~(\ref{eq:exppl}):  the red (dashed) line shows the exponential term and the green (solid) one - the power law.}
\end{figure}
\begin{figure}[h]
\includegraphics[width = 8cm]{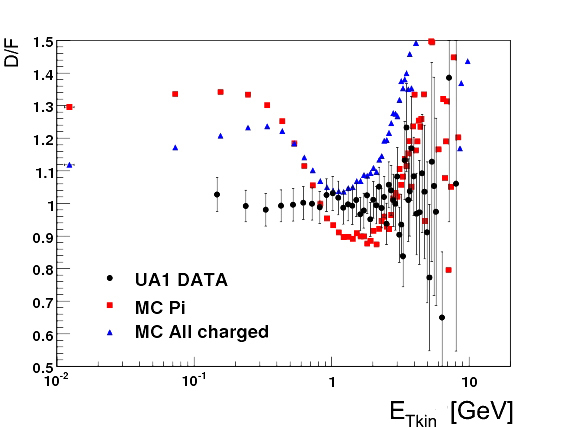}
\caption{\label{fig:05}Data-to-fit ratios for the experimental~\cite{UA1} and MC generated data divided by the same resulting fit function~(\ref{eq:exppl}) with the parameters obtained for the experimental spectrum.}
\end{figure}
\begin{figure}[h]
\includegraphics[width = 8cm]{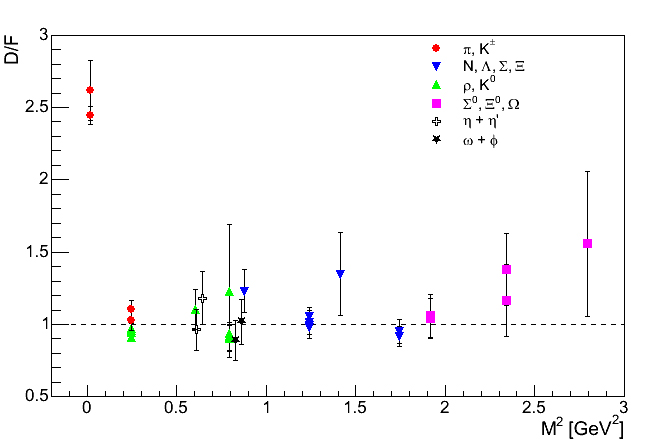}
\caption{\label{fig:06} The excess of the pion production rate over all other charged particles observed in hadronic decays of Z-bosons. The data are taken from~\cite{Chliapnikov}.}
\end{figure}

It is known that the vast majority of pions in hadronic collisions is produced via multiple cascade decays of the heavier  hadronic resonances. This might result in a transformation of some part of the initial power law spectrum of the produced short-lived heavy hadrons into an exponential decay pion distribution. These arguments could lead to an explanation of the observed anomaly in the pion production behavior. It is also hypothesized that charged kaons and baryons are more frequently produced directly in collisions than pions do. This hypothesis was tested using PYTHIA MC event generator~\cite{Sjostrand:2006za,Sjostrand:2007gs}. Despite the models used in the present event generators for direct hadron production in particle collisions are purely phenomenological, the cascade decay processes are described in these models quite accurately. The events were generated for high energy $pp-$collisions using PYTHIA 8.1 with a minimum bias set of parameters. The obtained MC final state pion spectrum shown in Fig~\ref{fig:04} is overlaid with results of the fitting of the function~(\ref{eq:exppl}) to the MC data. As it is seen from Fig~\ref{fig:04} the contribution of the exponential term to the final pion spectrum is minimal and does not exceed $15\%$. Moreover, this contribution dominates the spectrum at relatively high $P_T$ values around $2~GeV$. This MC behavior contradicts with the observations made before using the real hadronic collision data. Therefore, it looks very unlikely that the observed pion anomaly spectrum is due to the multiple cascade decays. In addition, it is worth mentioning that the overall reproduction of the shape of the experimental spectrum for charged particle (as well as pion) production in $pp-$collisions by MC event generator is quite poor. In Fig~\ref{fig:05} the ratio of the charged particle spectrum measured in the UA1 experiment~\cite{UA1} to the resulting fit function (D/F) is shown. As it is expected this ratio is close to unit within the experimental errors for the whole range of the measured charged particle transverse momentum.  The results of the MC calculations made for the UA1 experimental conditions divided by the same resulting fit function show strong dependence on $P_T$, thus the MC data have different spectrum shape than one observed in the experimental data.  


Finally, it is worth mentioning that pions have already demonstrated an anomalous behavior as concerns the absolute production rates with respect to all other hadrons. It was found~\cite{Chliapnikov}  that all hadrons produced in  hadronic decays of Z-bosons are produced with a probability described by an exponential function of the hadron mass squared. Only pions are produced about three times more frequently than it is expected from the extrapolation of this exponential function down to the pion mass value. A ratio of the measured hadron production rates to the exponential function approximating them is shown in Fig~\ref{fig:06}. This anomaly had not found any consistent explanation yet. Whether this anomaly of the pion production rate observed in the hadronic Z-boson decays and the anomaly of the pion production spectrum shape reported here are directly bounded to each other is unknown. Nevertheless, it is interesting to note that the size of the effect in both cases is about of the same order.


In conclusion the spectra shapes
of inclusive charged hadrons produced in high energy collisions were analyzed.
It was found that only pions require a sizable exponential term in their production spectra. This anomaly observed in pion production didn't find a consistent explanation yet.

The authors thank professors M.Ryskin and T.Sjostrand for helpful discussions.
This work was partially supported by Presidential grant for Scientific Schools in Russia.


\end{document}